\documentclass[aps,
prl,
twocolumn,
floatfix,
superscriptaddress,10pt
]{revtex4-2}
\usepackage{amssymb,amsmath,amsthm,graphicx,color}
\usepackage{bm}
\usepackage{mathrsfs}
\usepackage[bottom]{footmisc}
\usepackage{enumitem}
\usepackage{booktabs}
\usepackage{tabularx}
\usepackage{cancel}
\usepackage{pifont}
\usepackage{soul}
\usepackage[colorlinks]{hyperref}
\usepackage{mathtools,mathrsfs,esint}
\usepackage{array}
\usepackage{xcolor}
\usepackage{flushend}

\hypersetup{
     breaklinks=true,
    pdfstartview={FitH},    
    colorlinks=true,       
    linkcolor=blue,          
    citecolor=red,        
    filecolor=magenta,      
    urlcolor=blue,           
    anchorcolor=green,      
    linktocpage=true
}

\newcolumntype{C}[1]{>{\centering\arraybackslash}p{#1}}

\allowdisplaybreaks

\newcommand{\PRLsection}[1]{\emph{#1.---}}

\begin{document}

\title{{\bf Non-linear evolution of five-dimensional black strings in effective field theory}}

\author{Pau Figueras}
\email[]{p.figueras@qmul.ac.uk}
\author{\'Aron D. Kov\'acs}
\email[]{a.kovacs@qmul.ac.uk}
\author{Shunhui Yao}
\email[]{s.yao@qmul.ac.uk}

\affiliation{Centre for Geometry, Analysis and Gravitation, School of Mathematical Sciences, Queen Mary University of London, Mile End Road, London E1 4NS, United Kingdom}

\date{\today}

\begin{abstract}
We use numerical relativity to study the non-linear instability of five-dimensional black strings in Einstein--Gauss--Bonnet gravity. Black strings evolve into a series of black holes joined by thinner string-like segments, but key features of the dynamics depend on the sign of the Gauss–Bonnet coupling. For positive coupling, favored by UV considerations, the growth of curvature invariants is limited within the validity of effective field theory (EFT), suggesting a mechanism for restoring weak cosmic censorship. For negative coupling this cap is absent and curvatures may grow until the EFT breaks down.

\end{abstract}

\maketitle

\PRLsection{Introduction}%
The weak cosmic censorship (WCC) conjecture asserts that in general relativity (GR) the maximal future development of generic initial data should possess a complete null infinity \cite{Geroch:1979uc,Christodoulou_1999}. Essentially, this means that, generically, singularities should not be visible from infinity, i.e., they are ``hidden'' behind a horizon. While some version of the conjecture is believed to be true in $4$ spacetime dimensions, the status of WCC is drastically different for GR in dimensions greater than $4$ due to the Gregory-Laflamme (GL) instability \cite{Gregory:1993vy,Gregory:1994bj}. The GL instability was originally discovered as a linear instability of black string and $p$-brane solutions but more recent works observed that it is the underlying mechanism of the instability of black rings and even black holes with flat (AF), anti-de Sitter (AAdS) or Kaluza-Klein (KK) asymptotics \cite{Emparan:2001wn,Emparan:2003sy,Dias:2009iu,Dias:2010eu,Dias:2010gk,Santos:2015iua,Bantilan:2019bvf,Andrade:2020dgc}.

In a remarkable work, Lehner and Pretorius \cite{Lehner:2010pn} studied the fully non-linear evolution of the GL instability of black strings using numerical relativity, and found convincing evidence that a singularity would form in finite asymptotic time. Furthermore, they showed that the unstable black string evolves into a sequence of quasi-stationary spherical black holes connected by approximately uniform black strings. The black string segments are all GL unstable, so a cascade of GL instabilities develops on ever decreasing length scales, and hence increasingly fast timescales, which  results in the string reaching zero thickness (i.e., pinching off) in finite time. In GR horizons cannot smoothly bifurcate \cite{Hawking:1973uf} so a naked singularity has to appear at the pinch-off of the black string. This picture was confirmed and further refined in \cite{Figueras:2022zkg}, and extended to AF rotating black holes and black rings \cite{Figueras:2015hkb,Figueras:2017zwa,Bantilan:2019bvf}. While the actual details of the pinch-off of black strings are still unknown and the (in)completeness of future null infinity has not been fully addressed, these works provide generic examples of the observability of arbitrarily large curvatures that are not shielded by horizons.

However, GR should be understood as a low energy effective field theory (EFT) of gravity that is superseded by a UV complete theory of gravity (e.g., string theory). The corrections to GR due to UV physics become important exactly in regions of large curvature, and thus they are expected to have a significant impact on the non-linear dynamics of black strings. In this paper we take a UV-agnostic approach to this problem and employ the bottom-up EFT framework: we consider an EFT with a diffeomorphism-invariant action organised in a derivative expansion where the leading Einstein-Hilbert term is complemented by higher-curvature corrections. Two qualitatively different scenarios are possible:
\begin{itemize}
\item[(i)] The higher-derivative corrections \emph{regulate} the pinch-off, halting the cascade. In this case the endpoint is a regular configuration within the regime of validity of EFT and WCC is restored at low energies.
\item[(ii)] The higher-derivative corrections \emph{do not} stop the cascade before the thickness of the string becomes of the order of the UV length scale. In this case the EFT itself breaks down and the EFT is inconclusive about the fate of the endpoint of the GL instability.
\end{itemize}
Distinguishing between these scenarios for a concrete EFT of gravity is the main objective of this letter. More specifically, we consider the EFT of vacuum gravity in five spacetime dimensions, truncated at the leading order corrections to GR (thus neglecting Riemann$^3$ and higher-order operators). The resulting EFT is Einstein-Gauss-Bonnet (EGB) theory with the action
\begin{equation}
    S=\frac{1}{16\pi G}\int d^5x\sqrt{-g}\left(R+\lambda^\text{GB}\,\mathscr{L}_\text{GB}\right)\,, \label{eq:action5d}
\end{equation}
where $\lambda^\text{GB}$ is the coupling constant and $\mathscr{L}_\text{GB}$ is the Gauss-Bonnet invariant:
\begin{equation}
    \mathscr{L}_\text{GB}\equiv R^2 -4\,R_{\mu\nu}\,R^{\mu\nu}+R_{\mu\nu\rho\sigma}\,R^{\mu\nu\rho\sigma}\,.
\end{equation}
The coupling constant $\lambda^\text{GB}$ has dimensions of length$^2$, hence $\sqrt{|\lambda^\text{GB}|}$ sets the length scale of new UV physics. The equations of motion of the theory are given by
\begin{equation}
    G^{\mu}{}_{\nu} + \lambda^\text{GB}\,\delta_{\nu \rho_1 \rho_2 \rho_3 \rho_4}^{\mu \sigma_1 \sigma_2 \sigma_3 \sigma_4} R_{\sigma_1\sigma_2}{}^{\rho_1 \rho_2}R_{\sigma_3\sigma_4}{}^{\rho_3 \rho_4}=0, \label{eq:eomsEGB}
\end{equation} 
where $\delta^{\mu_1\cdots\mu_q}_{\nu_1\dots\nu_q}\equiv q!\delta^{\mu_1}_{[\nu_1}\cdots\delta^{\mu_q}_{\nu_q]}$ is the generalized Kronecker delta. The equations of motion \eqref{eq:eomsEGB} are second order in derivatives and the theory admits a well-posed initial value formulation in a modified harmonic gauge \cite{Kovacs:2020ywu} and in singularity-avoiding coordinates \cite{AresteSalo:2023hcp} (see also \cite{Figueras:2024bba} for an alternative formulation of the initial value problem for a general class of EFTs). While these works establish EGB theory as a consistent classical EFT for either sign of $\lambda^\text{GB}$, we mention that UV considerations impose constraints on the GB coupling. These constraints follow from the standard dispersive (positivity) argument applied to $2\to2$ graviton scattering: analyticity, crossing, and unitarity imply that in any tree-level UV completion of gravity that is free of ghosts and tachyons, $\lambda^\text{GB}$ must be non-negative in dimensions $D > 4$~\cite{Cheung:2016wjt,Bellazzini:2015cra}. This positivity constraint is consistent with the sign of the Gauss-Bonnet coupling in the low-energy effective action of heterotic and bosonic string theory \cite{Zwiebach:1985uq,Gross:1986mw,Gross:1986iv,Metsaev:1987zx,Boulware:1985wk}. On the other hand, inconsistent applications of strongly coupled EGB theory (i.e., outside the EFT regime of validity) are known to lead to pathologies, see e.g., \cite{Camanho:2014apa,Papallo:2015rna,Chen:2021bvg}.

\PRLsection{Conventions} We use Greek letters $\alpha,\beta,\ldots$ to denote spacetime coordinates. The first letters of the Latin alphabet, $a,\,b\,,\ldots \in\{t,x,z\}$ are used to denote the directions transverse to the round $S^2$ of the black strings, $w_1,w_2, \ldots$ are coordinate indices on the sphere. We use the curvature conventions of \cite{Wald:1984rg}.

\PRLsection{Setup and numerical methods} We restrict to spherical symmetry along the AF directions, which reduces the dimensionality of the problem whilst capturing the essential physics. We introduce coordinates $(z,w_1,w_2)$ through $z=r\cos\theta$, $w_1=r\sin\theta\cos\phi$, $w_2=r\sin\theta\sin\phi$ and implement the modified cartoon reduction of \cite{Pretorius:2004jg,Yoshino:2009xp,Cook:2016soy} by restricting the system to the slice $w_1=w_2=0$, $z\ge0$. We perform the numerical simulations with an extended version of the \texttt{GRChombo} code previously used in \cite{Figueras:2022zkg}. Our code implements  the evolution equations of EGB theory in the mCCZ4 formulation \cite{AresteSalo:2022hua,AresteSalo:2023mmd} using the core features of \texttt{GRFolres} \cite{AresteSalo:2023hcp}. The latter code has been extensively used in numerical simulations of four-dimensional Einstein–scalar–Gauss–Bonnet gravity \cite{AresteSalo:2023mmd,Doneva:2023oww,Doneva:2024ntw,AresteSalo:2025sxc,Corman:2025wun}.

\begin{figure}[t!]
    \centering
    \includegraphics[width=\linewidth]{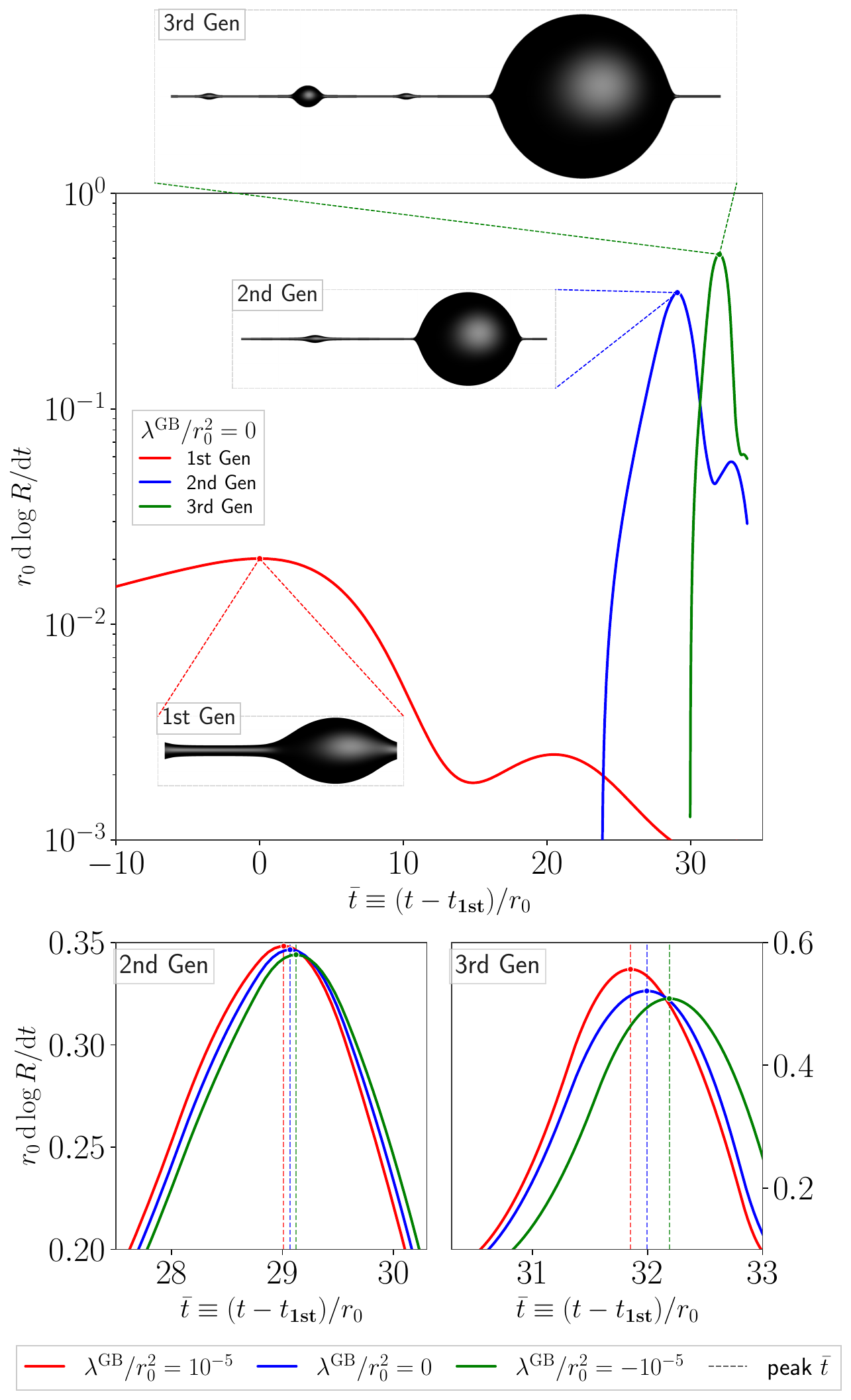}
    \caption{Growth-rate profile of the first three generations of bulges for different couplings,
    plotted against the shifted time $\bar t=(t-t_{\rm 1st})/r_0$, with $t_{\rm 1st}$ the first-generation peak time for each run.
\textit{Top}: Three-generation profile of $r_0\,d\log R/dt$ in the GR evolution, with the corresponding apparent horizon embedding diagrams displayed in the insets.
\textit{Bottom}: Zoomed-in views of the second (left) and third (right) generation peaks for $\lambda^{\mathrm{GB}}/r_0^2=\{10^{-5},0,-10^{-5}\}$.
The dashed vertical lines mark the extracted peak times. 
For $\lambda^\text{GB}>0$, the higher generations form earlier than in GR, while for $\lambda^\text{GB}<0$, they form later.  }
    \label{fig:time}
\end{figure}

For a clean comparison with \cite{Figueras:2022zkg}, we use the same initial conditions \footnote{Note that these initial conditions do not solve the EGB constraint equations. To minimise constraint violations, we initially set $\lambda^\text{GB}$ to zero and gradually switch it on as a quadratic function of time until it reaches its target value, which we keep fixed for the rest of simulations.}, constraint-damping parameters, and mesh-refinement scheme (see Sec.II of \cite{Figueras:2022zkg} for more details). We fix the length of the periodic direction to $L = 10$ and the initial black string radius to $r_0 = 1$. The dimensionless parameter $\lambda^\text{GB}/r_H^2$, where $r_H$ denotes the local apparent horizon (AH) radius along the black string,  provides a measure of the strength of the GB coupling. This ratio is chosen to be small initially but gradually increases during the non-linear evolution. To mitigate the singular behaviour of higher-derivative terms near the centre of the black string, we follow \cite{AresteSalo:2023mmd} and smoothly switch off the Gauss--Bonnet coupling  well inside the AH, ensuring that this regularisation does not affect the exterior dynamics \footnote{More specifically, we do this by replacing $\lambda^{\mathrm{GB}} \rightarrow \frac{\lambda^{\mathrm{GB}}}{1+e^{-100(\chi-\chi_0)}}$ with $\chi_0 = 0.1$ and $\chi$ denotes the (m)CCZ4 conformal factor. The $\chi = 0.1$ level set lies well inside the AH \cite{Andrade:2020dgc,Figueras:2022zkg}.}.

We consider couplings of both signs, $\lambda^{\mathrm{GB}}/r_0^2\in\{\pm10^{-4},\pm10^{-5}\}$ and also perform GR simulations ($\lambda^\text{GB}=0$) as benchmarks. The mCCZ4 gauge functions are chosen as (i) $a(x)=0$, $b(x)=0.1$ and (ii) $a(x)=0.001$, $b(x)=0.4$ 
\footnote{
We employ choice (i) for the $\lambda^{\mathrm{GB}}/r_0^2 = 0,\pm10^{-5}$ simulations and choice (ii) for the $\lambda^{\mathrm{GB}}/r_0^2 = 10^{-4}$ simulations, because the run with the larger coupling generally requires a more substantial modification in the propagation speeds of the unphysical modes. 
Note that while $a(x)$ is generally taken to be non-zero \cite{AresteSalo:2023mmd}, in the symmetry-reduced setting of this work the choice $a(x)=0$ also leads to a strongly hyperbolic evolution system.}.

\PRLsection{Coupling dependence of apparent horizon dynamics} The non-linear evolution of the GL instability in EGB theory is qualitatively similar to GR \cite{Lehner:2010pn,Figueras:2022zkg}: a cascade of satellite black holes connected by thinner string segments whose thickness decreases over time. Following \cite{Figueras:2022zkg}, we track the relative growth rate $r_0 R_h^{-1}\dot{R_h}\equiv r_0 d\ln R_h/dt$ of the horizon radii $R_h$ at selected local bulges. This quantity exhibits a repetitive pattern: for each generation, the growth rate rises to a peak and then decreases to a trough (see the upper panel of Fig.~\ref{fig:time}), allowing us to identify the peak (and trough) times $t_{p,i}$ ($t_{n,i}$) unambiguously. In order to remove the possible sensitivity to initial transients, we measure the time relative to the first generation peak time, $ \bar t \equiv (t-t_{\rm 1st})/r_0$. We observe in EGB theory that once the first generation has formed, the subsequent cascade is governed by a universal dynamics independently of its earlier history, similarly to GR \cite{Lehner:2010pn}.

\begin{figure}[t!]
    \centering
    \includegraphics[width=\linewidth]{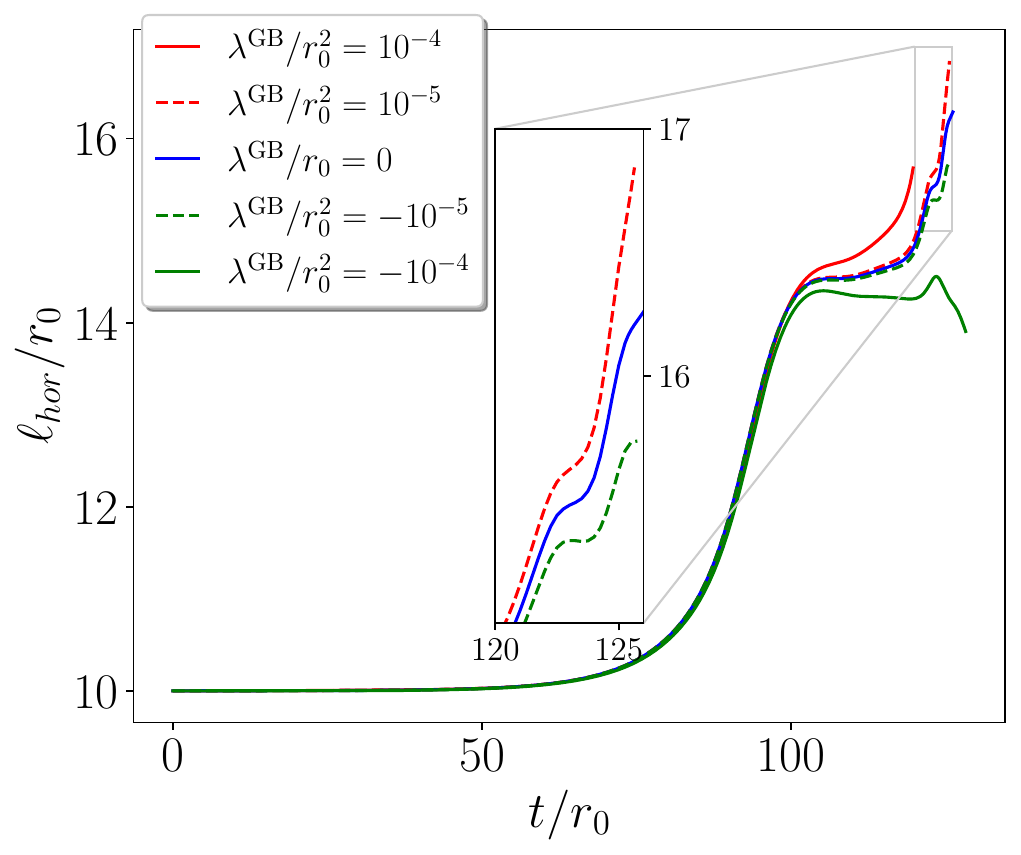}
    \caption{Time evolution of proper length of black string for different couplings. For $\lambda^\text{GB}>0$, the local tension of the string is lower than in GR, and hence the proper length increases during the evolution. For $\lambda^\text{GB}<0$, the tension is higher than in GR and hence strings become shorter and even contract. }
    \label{fig:proplen}
\end{figure}

Figure \ref{fig:time} illustrates the formation dynamics of the first three generations of black holes in terms of the relative growth rate of the radii. In the upper panel, we only display the $\lambda^{\mathrm{GB}}/r_0^2=0$ result, and we provide zoomed-in views of the second- and third generation peaks for various couplings in the lower panel. We observe a monotonic phase shift in the formation of black hole generations with respect to the sign of the coupling:  $\lambda^{\mathrm{GB}}>0$ moves the peaks forward compared to the GR run (i.e., the blobs form faster), whereas $\lambda^{\mathrm{GB}}<0$ brings them to later $\bar t$ (i.e., the blobs form more slowly). Thus, the Gauss-Bonnet correction accelerates the GL instability cascade for $\lambda^{\mathrm{GB}}>0$, while for $\lambda^{\mathrm{GB}}<0$ the cascade slows down.

We also track the proper length $\ell_{hor}$ of the AH along the string which gives a direct geometric measure of the stretching of the black string. Recall that in GR, the blob/string dynamics implies that  the shape of the  AH  develops a fractal structure which causes $\ell_{hor}$ to diverge at the pinch off. In Fig.~\ref{fig:proplen}, we show the proper length $\ell_{hor}$ versus time for different coupling values. At late times, the profiles of these curves depend significantly on the sign of the coupling as the accumulated effect of the Gauss–Bonnet corrections becomes important. In particular, the larger the value of $|\lambda^\text{GB}|/r_0^2$, the earlier the deviation from the GR behaviour occurs. For $\lambda^{\mathrm{GB}}>0$, the proper length of the AH grows faster than in GR, indicating that the Gauss-Bonnet term enhances the late time stretching of the horizon. For $\lambda^{\mathrm{GB}}<0$, the growth of the proper length slows down. This is particularly visible for $\lambda^{\rm GB}/r_0^2=-10^{-4}$, where the proper length even decreases at late times.

This behaviour is consistent with the intuition gained from the study of the physical properties of uniform black strings (UBS) and their linear perturbations (see the analysis of our companion paper \cite{Figueras:2026evh} and also \cite{Kobayashi:2004hq,Suranyi:2008wc,Brihaye:2010me} for details). The onset of the GL instabilities is controlled by the thickness of the black string: longer and thinner string segments are more prone to developing the next generation of instability. The faster (resp. slower) growth of $\ell_{\rm hor}$ for $\lambda^{\rm{GB}}>0$ (resp. $\lambda^{\rm{GB}}<0$) will therefore result in faster (resp. slower) formation of the higher-generation satellites. This is related to the fact UBSs with $\lambda^{\rm{GB}}>0$ have lower tension than their counterparts in GR and hence, intuitively, it is ``easier'' for them to develop non-uniformities along the string direction. On the other hand, for $\lambda^{\rm{GB}}<0$, UBSs in EGB have larger tension than in GR.   However, we emphasise that this picture is not accurate in the $\lambda^{\rm{GB}}>0$ case at late times when the segments joining the blobs become highly non-uniform and are no longer captured by quasi-stationary UBS solutions as we shall explain shortly.

\PRLsection{Endpoint of the GL instability of black strings in EGB gravity}%
 Although the results of our simulations provide evidence for two distinct scenarios depending on the sign of the GB coupling (scenario (i) of the Introduction for positive and scenario (ii) for negative couplings), it seems that a definitive resolution of the end state is only possible in a particular UV completion of GR.

For all $\lambda^{\mathrm{GB}}\neq 0$, our simulations eventually abort when the string becomes sufficiently thin (of the order of the UV cutoff scale). This is due to the change of character of the effective metric that controls the propagation of the physical degrees of freedom (DOF).  In $5$ dimensions, a theory of gravity with second order equations of motion (such as GR or EGB theory) admits $5$ physical DOF. However, in the symmetry-reduced setting studied in this Letter, only $1$ DOF is excited. As discussed in our companion paper \cite{Figueras:2026evh} (see also \cite{Reall:2021voz,Thaalba:2024crk,R:2022hlf}), the dynamics and causal structure of EGB theory in KK spaces with spherical symmetry along the AF directions are determined by the effective metric \footnote{The constant factor $\tfrac{16}{3}$ is chosen such that $g_{\textbf{eff}}^{ab}$ reduces to the spacetime metric $g^{ab}$ as $\lambda^{\mathrm{GB}}\to0$.}
\begin{equation}\label{eq:g_eff}
     g_{\textbf{eff}}^{ab}=\tfrac{16}{3}g^{kb}\delta^{ace}_{kdf}(8\,N^{dw}_{cw}N^{fw}_{ew}-N^{df}_{ce}N^{w_1w_2}_{w_1w_2}),
\end{equation}
where
\begin{equation}
    N^{\mu\nu}_{\rho\sigma}=\tfrac{1}{8}\bigl(\delta^{\mu\nu}_{\rho\sigma}+4\,\lambda^{\mathrm{GB}}\,R^{\mu\nu}_{\rho\sigma}\bigr).
\end{equation}
At weak coupling, $g_{\textbf{eff}}$ has Lorentzian signature, a necessary condition for a well-posed initial value formulation of EGB theory. However, when $\lambda^{\mathrm{GB}}|R^{\mu\nu}_{\rho\sigma}|$ becomes of order $1$, $g_{\textbf{eff}}$ may change signature, resulting in the EGB equations of motion changing character from hyperbolic to elliptic and the breakdown of Cauchy evolution, \emph{regardless of gauge choice}. In the End Matter and in our companion paper \cite{Figueras:2026evh}, we provide more details for specialist readers on detecting the loss of hyperbolicity.

\begin{figure}[t!]
    \centering
    \includegraphics[width=\linewidth]{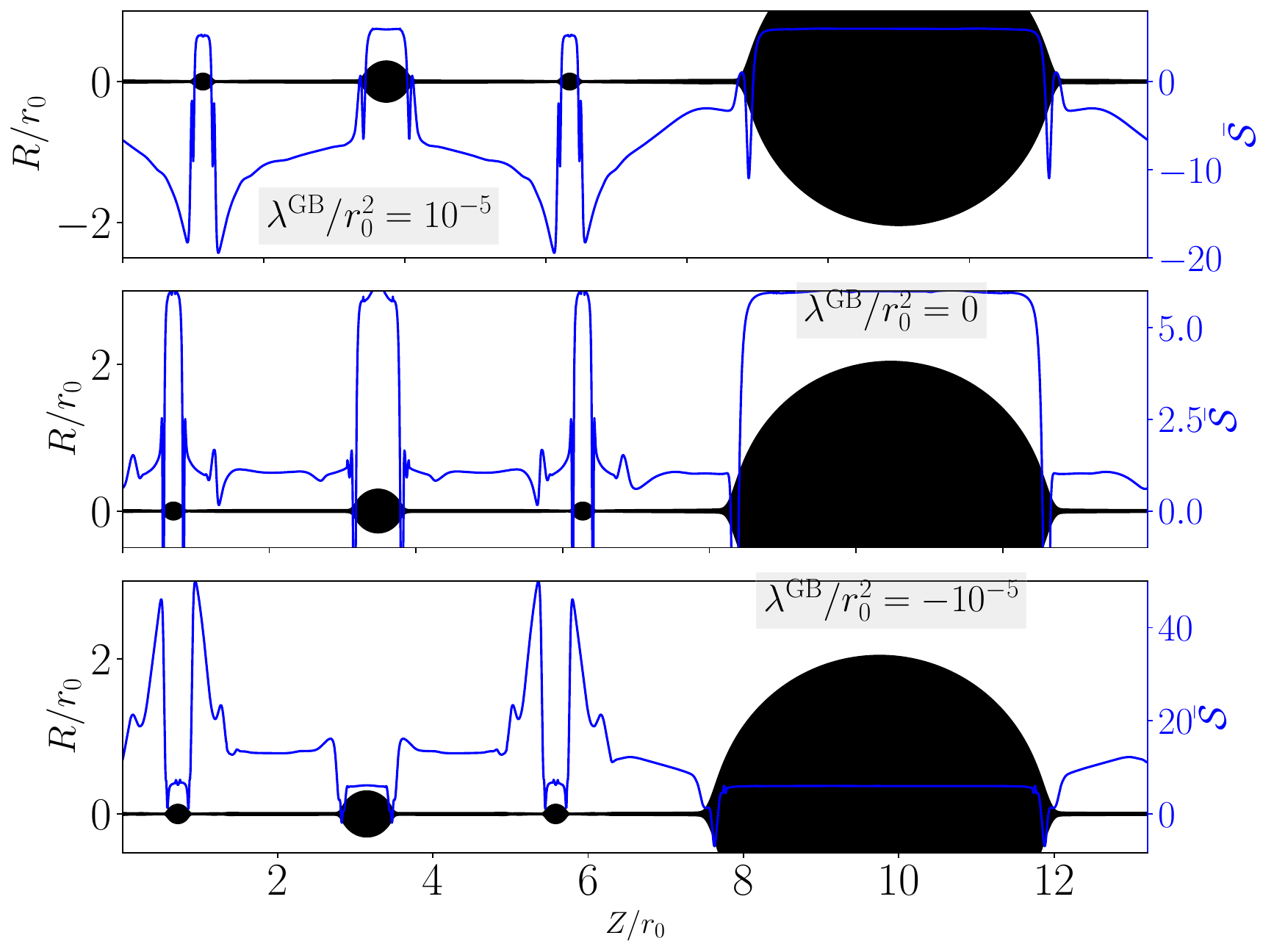}
    \caption{The cubic invariant $\bar{\mathcal{S}}$ on the AH, evaluated at the last stage of the evolutions, for $\lambda^{\mathrm{GB}}/r_0^2=\{10^{-5},\,0,\,-10^{-5}\}$ from top to bottom.}
    \label{fig:rime3}
\end{figure}

To describe the end state of the strings, in Fig.\,\ref{fig:rime3} we plot the normalized cubic invariant $\bar{\mathcal{S}}=~324\,J^2\,I^{-3} + 26$ on the AH in the last snapshot of our simulations for $\lambda^{\mathrm{GB}}/r_0^2={10^{-5}},\,0,\,-10^{-5}$, from top to bottom. Here $I\equiv R_{\mu\nu\rho\sigma}\,R^{\mu\nu\rho\sigma}$ is the Kretschmann invariant and $J\equiv R^{\mu\nu}{}_{\rho\sigma}\,R^{\rho\sigma}{}_{\alpha\beta}\,R^{\alpha\beta}{}_{\mu\nu}$. The normalization $\bar{\mathcal{S}}$ has been chosen so that this quantity is equal to $1$ on the AH of a UBS in GR. 

The profile of $\bar{\mathcal{S}}$ along the string shows significant differences for the different sign choices of $\lambda^\text{GB}$. For $\lambda^\text{GB}<0$ (bottom panel of Fig.\,\ref{fig:rime3}), the behaviour of $\bar{\mathcal{S}}$ is qualitatively similar to the GR case: the string segments that join the bulges are fairly uniform. This is demonstrated by the fact that the cubic invariant $\bar{\mathcal{S}}$ remains positive and approximately constant along these segments. For $\lambda^\text{GB}<0$, our companion paper \cite{Figueras:2026evh} shows that UBSs have a lower bound on their thickness but the corresponding limiting solution has diverging curvature invariants on the AH. Moreover, $g_{\textbf{eff}}^{ab}$ has Lorentzian signature on the horizon and in the exterior of UBS backgrounds in this case. Therefore, there seems to be no mechanism limiting the growth of curvature invariants in the non-linear evolution, allowing the magnitude of $\bar{\mathcal{S}}$ to grow significantly more in our simulations than in the $\lambda^\text{GB}>0$ case. On the other hand, for $\lambda^\text{GB}>0$ (top panel in Fig.\,\ref{fig:rime3}) the joining string segments become non-uniform, shown by the non-constant profile of $\bar{\mathcal{S}}$. Note also that $\bar{\mathcal{S}}$ takes negative values along the string as opposed to the $\lambda^\text{GB}\leq 0$ cases. The thickness of the string segments for $\lambda^\text{GB}>0$ approaches (but remains above) the value $\sqrt{4\lambda^\text{GB}/(\sqrt2 -1)}\approx 3.1\sqrt{\lambda^\text{GB}}$ set by the criterion that $g_{\textbf{eff}}^{ab}$ remains Lorentzian on and outside the horizon of UBS backgrounds (see \cite{Figueras:2026evh}), which is above the minimum thickness $\sqrt{8\,\lambda^\text{GB}}$ of UBSs in EGB theory with $\lambda^\text{GB}>0$. Crucially, all UBSs with $r_H\geq \sqrt{8\lambda^\text{GB}}$ have finite curvature on and outside of their horizons. This mechanism (which is somewhat similar to strong cosmic censorship \cite{Reall:2014pwa,Ripley:2019hxt}) appears to prevent the extension of Cauchy evolution further. This means that thinner black strings may be dynamically unattainable, placing a (large but finite) upper bound on the growth of curvature invariants and indicating that the pinch-off might be censored in a fundamental theory with $\lambda^\text{GB}>0$.

\begin{figure}[t!]
    \centering
    \includegraphics[width=\linewidth]{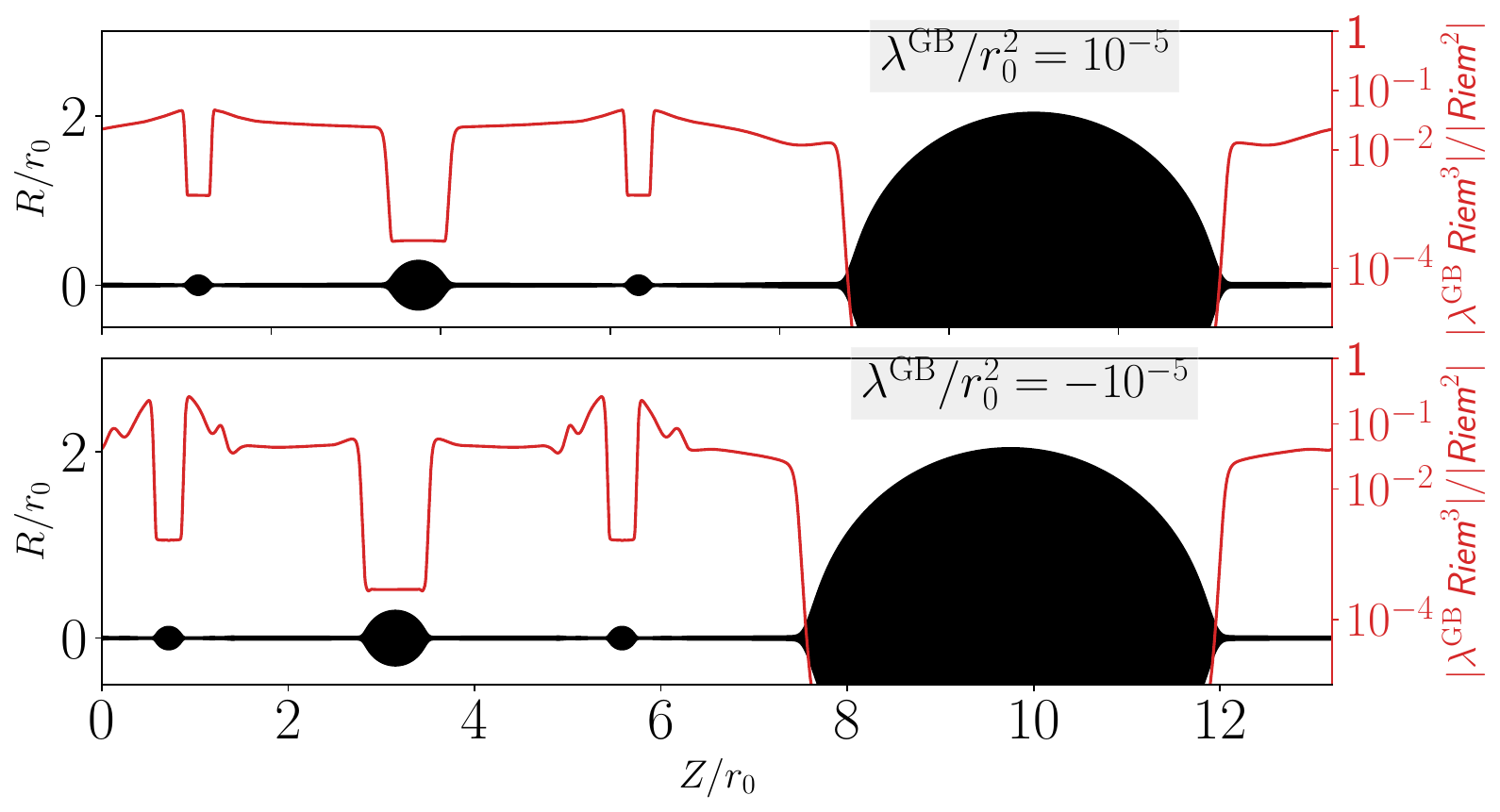}
    \caption{The derivative expansion condition $|\lambda^{\mathrm{GB}} \mathbf{Riem}^3|\allowbreak/|\mathbf{Riem}^2|$ on the AH, evaluated at the last stage of the evolutions for $\lambda^{\mathrm{GB}}/r_0^2=\{10^{-5},-10^{-5}\}$ from top to bottom.}
    \label{fig:derivative}
\end{figure}

One may wonder if the breakdown of hyperbolicity is related to the breakdown of the derivative expansion in \eqref{eq:action5d}. To estimate the impact of the next-to-leading order term neglected in \eqref{eq:action5d}, we evaluate the dimensionless ratio $|\lambda^{\mathrm{GB}} \mathbf{Riem}^3|/|\mathbf{Riem}^2|$ on our solutions. In Fig.\,\ref{fig:derivative} we display this ratio on the AH of the string at the end of our simulations. We can see that for $\lambda^\text{GB}>0$, the maximum of $|\lambda^{\mathrm{GB}} \mathbf{Riem}^3|/|\mathbf{Riem}^2|$ is less than $0.1$ so the dynamics is still dominated by EGB theory, whereas for $\lambda^\text{GB}<0$, the maximum of $|\lambda^{\mathrm{GB}} \mathbf{Riem}^3|/|\mathbf{Riem}^2|$ is between $0.1$ and $1$, indicating that the six-derivative terms in \eqref{eq:action5d} become important and the EFT breaks down at the end of our simulation.

\PRLsection{Discussion}%
We have studied the non-linear GL instability of black strings in EGB gravity in five spacetime dimensions. Our main finding is that several details of the dynamics and the endpoint of the instability depend on the sign of the coupling $\lambda^\text{GB}$. For $\lambda^\text{GB}>0$, and hence for an EFT compatible with UV consistency conditions, we suggest a mechanism (akin to strong cosmic censorship) that caps the contraction of the string and the growth of curvature, thus potentially preventing a pinch-off and the formation of a naked singularity in finite time. Furthermore, in this case, higher derivative terms not included in \eqref{eq:action5d} remain subdominant compared to the Gauss-Bonnet term. On the other hand, for $\lambda^\text{GB}<0$, the dynamics is more similar to GR: although UBSs admit a minimum thickness, the limiting solution is singular, and the curvature invariants on and outside the AH may reach significantly larger values before the breakdown of EFT occurs.

Given that our simulations stop due to loss of hyperbolicity within the AH for both sign choices of $\lambda^\text{GB}$, it is natural to ask whether it is possible to extend the evolution using the regularisation method of \cite{Figueras:2024bba} to approach the maximum Cauchy development more closely in both cases \footnote{This would require a more aggressive strategy for switching off the Gauss-Bonnet coupling in the interior or a reliable excision.}. We will explore this in future work.
Although the EFT description seems to be inconclusive about the endpoint of the instability, for $\lambda^\text{GB}>0$ the capping mechanism mentioned above appears to halt the evolution when the AH radius reaches the UV length scale. From this point onwards, a Horowitz-Polchinski-type of transition \cite{Susskind:1993ws,Horowitz:1996nw,Horowitz:1997jc} could implement the topology change as advocated in \cite{Emparan:2024mbp}.

\section*{Acknowledgements}
We thank Katy Clough, Daniela Cors, Roberto Emparan and Luis Lehner for helpful discussions; and Llibert Aresté Saló, Tiago França and Chenxia Gu for technical support throughout the code development and numerical simulations. 
PF also wants to thank Mikel S\'anchez Garitaonandia, Savdeep Sethi and Robert Wald for further discussions. PF would also like to thank the Enrico Fermi Institute and Department of Physics of the University of Chicago, where a significant part of this work was carried out, for their hospitality. We also thank the GRTL collaboration (\texttt{www.grtlcollaboration.org}) for their support and code development work. PF and ADK are supported by the STFC Consolidated Grant ST/X000931/1  (Astronomy at Queen Mary 2023-2026). SY was supported by the China Scholarship Council.  The authors acknowledge the use of the Sulis HPC facility and the support of the Sulis Research Software Engineer (RSE) team. Sulis is funded by the Engineering and Physical Sciences Research Council (EPSRC) (grant number EP/T022108/1) and the HPC Midlands+ consortium. This work also used the DiRAC@Durham facility managed by the Institute for Computational Cosmology on behalf of the STFC DiRAC HPC Facility (www.dirac.ac.uk). The equipment was funded by BEIS capital funding via STFC capital grants ST/P002293/1, ST/R002371/1 and ST/S002502/1, Durham University and STFC operations grant ST/R000832/1. DiRAC is part of the National e-Infrastructure. We also acknowledge the EuroHPC Joint Undertaking for awarding this project access to the EuroHPC supercomputer LUMI-C, hosted by CSC (Finland) and the LUMI consortium through a EuroHPC Regular Access call grant number EHPC-REG-2025R01-217. Finally, the Texas Advanced Computing Center (TACC) at The University of Texas at Austin also provided computational resources that have contributed to the research results reported in this paper. URL: http://www.tacc.utexas.edu

\bibliography{reference}

\clearpage

\onecolumngrid
\begin{center}
\textbf{\large End Matter}
\end{center}
\vspace{0.5\baselineskip}
\twocolumngrid

\PRLsection{Geometric properties of the generations}
In Table \ref{tab:summary} we summarize some geometric properties of the bulges and connecting string segments for the first three generations and couplings $\lambda^\text{GB}/r_0^2=10^{-5},\,0,\,-10^{-5}$. For the first two generations, the differences between the various cases are small. This is because not enough time has elapsed for accumulated small effects of the weakly coupled EGB theory to make a big difference. However, the effects due to the EFT corrections become more pronounced by the time the third generation forms since curvatures on the AH have grown significantly and hence the Gauss-Bonnet term has a greater impact on the evolution. For $\lambda^{\mathrm{GB}}>0$ the descendent string segments are slightly thinner than in GR, while for $\lambda^{\mathrm{GB}}<0$ they are thicker. This is because strings have lower tension than in GR in the former case, while they have larger tension in the latter case.  The geometric properties of the bulges are less sensitive to the sign of the coupling because the local effective coupling $\lambda^\text{GB}/R_{h,f}^2$ on the blobs  is small, even for the higher generations.

\begin{table}[t]
    \centering
\begin{ruledtabular}
\begin{tabular}{c|ccccccc}
Gen. & $\lambda^{\mathrm{GB}}/r_0^2$ & $\bar{t}_{p,i}$ & $\bar{t}_{n,i}$ & $n_s$ & $R_{s,i}/r_0$ & $R_{h,f}/r_0$ & $R_{s,i}/L_{s,i}$ \\
\midrule
1 & $10^{-5}$ & 0 & $14.76$ & 1 & 1 & $2.038$ & 0.1 \\
 & $0$ & 0 & $14.82$ & 1 & 1 & $2.039$ & 0.1 \\
 & $-10^{-5}$ & 0 & $14.75$ & 1 & 1 & $2.037$ & 0.1 \\
\hline
2 & $10^{-5}$ & $29.01$ & $29.01$ & 1 & $0.1094$ & $0.2892$ & $0.01705$ \\
 & $0$ & $29.07$ & $31.67$ & 1 & $0.1097$ & $0.2957$ & $0.01715$ \\
 & $-10^{-5}$ & $29.12$ & $31.67$ & 1 & $0.11$ & $0.2832$ & $0.01725$ \\
\hline
3 & $10^{-5}$ & $31.85$ & N/A & 2 & $0.04456$ & $0.1162$ & $0.01321$ \\
 & $0$ & $31.99$ & N/A & 2 & $0.0454$ & $0.1192$ & $0.0139$ \\
 & $-10^{-5}$ & $32.19$ & N/A & 2 & $0.04627$ & $0.1118$ & $0.01458$ \\
\end{tabular}
\end{ruledtabular}
    \caption{A summary of the geometric properties of each generation of bulges and connecting string segments for $\lambda^\text{GB}/r_0^2=\{10^{-5},\,0,\,-10^{-5}\}$.}
    \label{tab:summary}
\end{table}

\begin{figure}[t!]
    \centering
    \includegraphics[width=\linewidth]{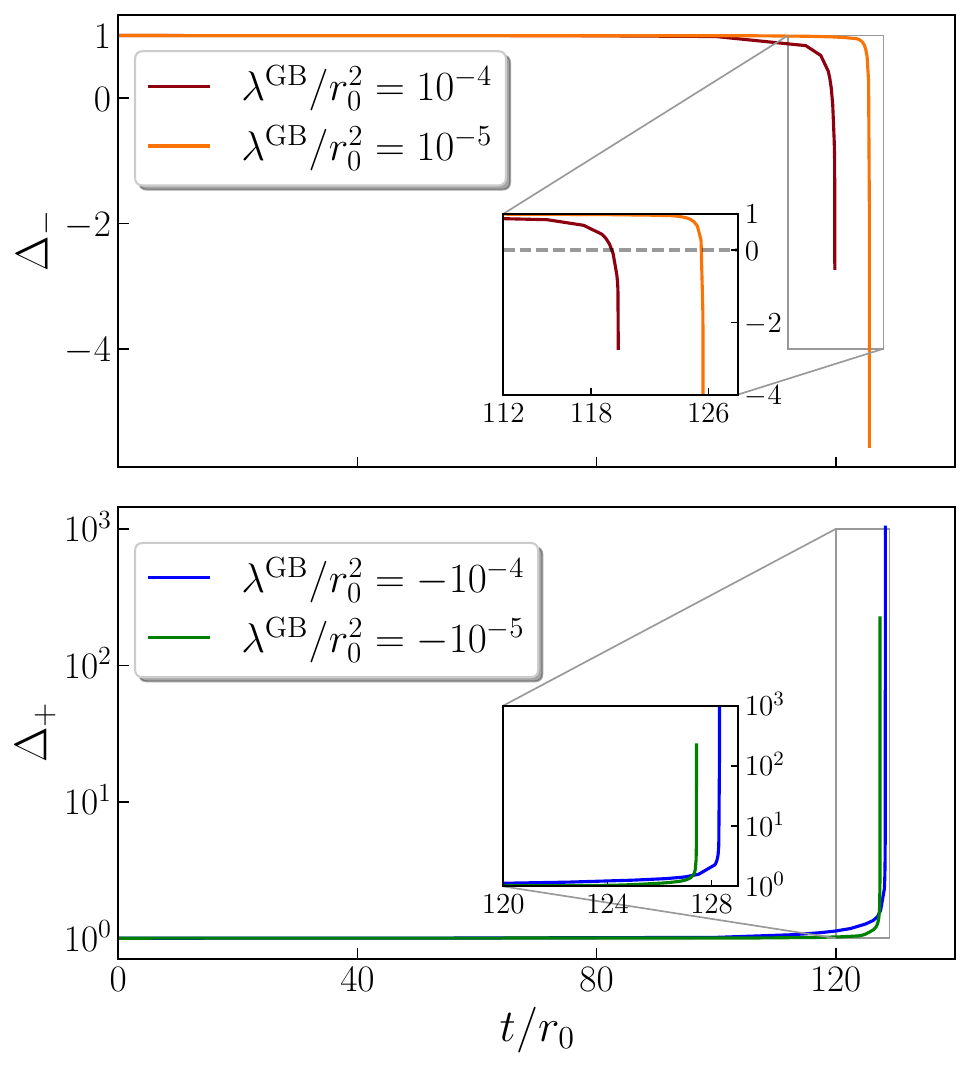}
    \caption{Signature of hyperbolicity loss. \textit{Top}: evolution of the global minimum of $\Delta_{-}$ for $\lambda^{\mathrm{GB}}/r_0^2=\{10^{-4},10^{-5}\}$. \textit{Bottom}: evolution of the global maximum of $\Delta_{+}$ for $\lambda^{\mathrm{GB}}/r_0^2=\{-10^{-4},-10^{-5}\}$.}
    \label{fig:hyper}
\end{figure}

\begin{figure}[t!]
    \centering
    \includegraphics[width=\linewidth]{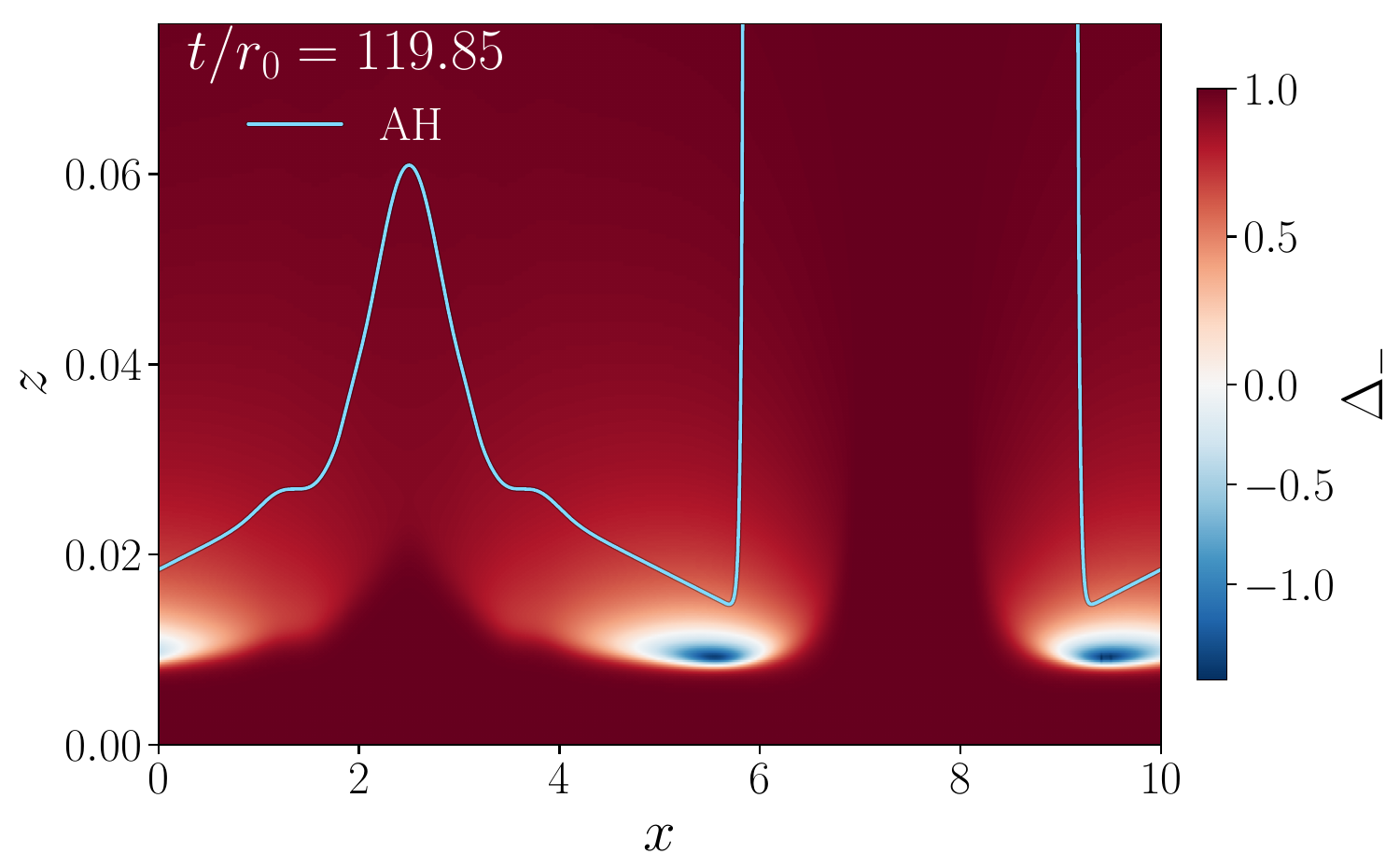}
    \includegraphics[width=\linewidth]{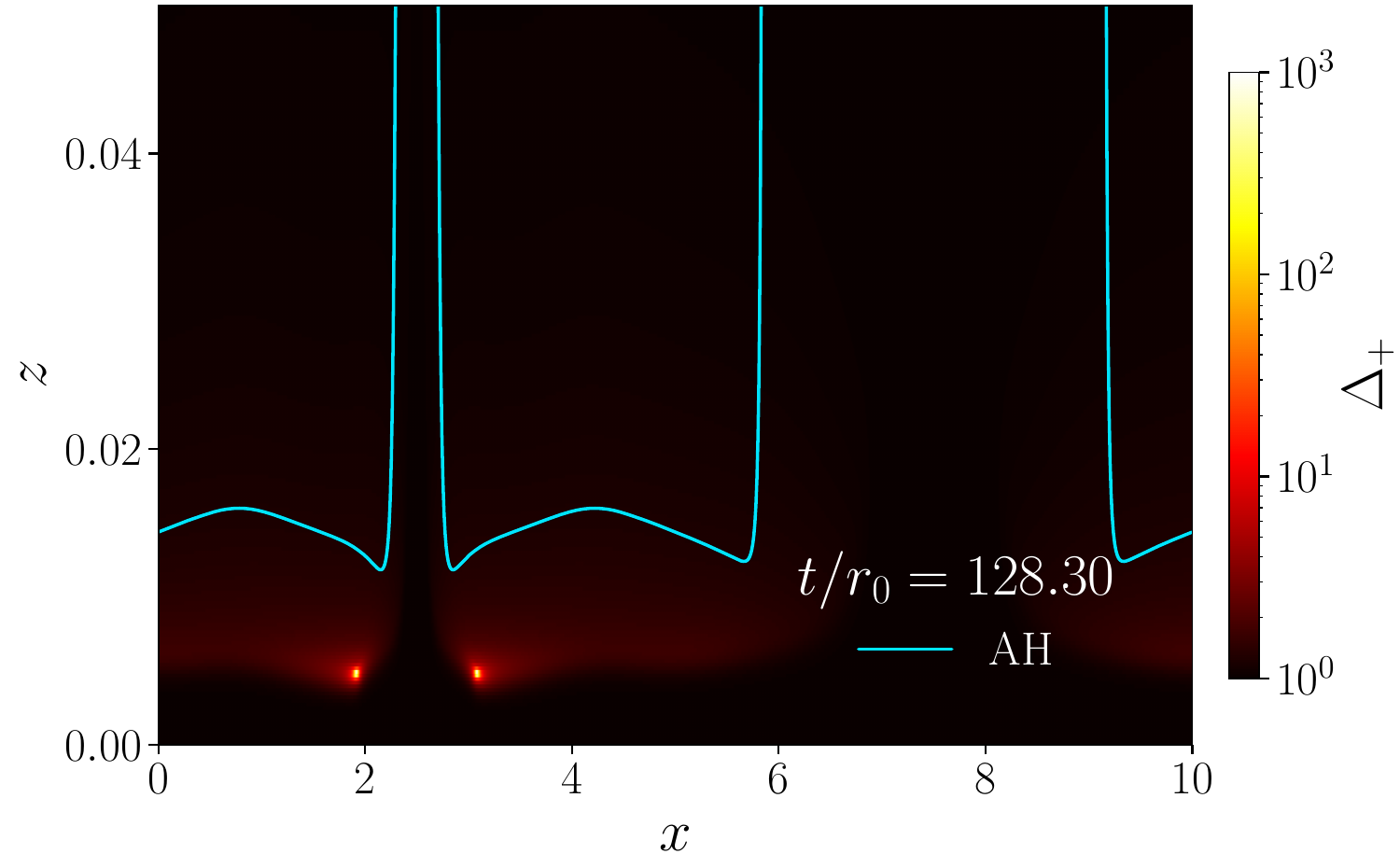}
    \caption{The spatial distribution of the diagnostics $\Delta_{-}$ (\textit{Top}) and $\Delta_{+}$ (\textit{Bottom}) in the final snapshot before the breakdown of the $\lambda^{\mathrm{GB}}/r_0^2=10^{-4}$ (\textit{Top}) and $\lambda^{\mathrm{GB}}/r_0^2=-10^{-4}$ (\textit{Bottom}) runs. The light blue line corresponds to the AH. These plots show that in both cases the transition occurs inside the AH, in the thinnest region of the string, where the neck joins one of the blobs.}
    \label{fig:AH_Delta}
\end{figure}

\PRLsection{Signatures of the hyperbolicity loss}
The change of character of the evolution equations can be identified using the signature change of the effective metric \eqref{eq:g_eff}. To diagnose this in our simulations, we found it convenient to introduce the induced effective metric $\Delta^{ij}$ by
\begin{equation}
    \Delta^{ij}\equiv -\frac{g_{\textbf{eff}}^{ij}}{g_{\textbf{eff}}^{00}}+\frac{g_{\textbf{eff}}^{0i} ~g_{\textbf{eff}}^{0j}}{(g_{\textbf{eff}}^{00})^2}\,,
\end{equation} 
together with its generalized eigenvalues
\begin{equation}
     \Delta^{ij}\xi_j=\Delta_{\pm}\Delta^{ij}_{\mathbf{gr}}\xi_j \implies \det(\Delta^{ij}-\Delta_{\pm}\Delta^{ij}_{\mathbf{gr}})=0,
\end{equation}
where the reference metric $\Delta_{\mathbf{gr}}^{ij}$ corresponds to the value of $\Delta^{ij}$ in the GR limit ($\lambda^\text{GR}=0$); and $\Delta_{+}$ and $\Delta_{-}$ denote the largest and the smallest generalized eigenvalues, respectively. By construction, $\Delta_\pm$ are invariant under spatial coordinate changes.
We  use the diagnostic quantities $\Delta_\pm$ to characterize the loss of hyperbolicity that causes the halting of our simulations. We can distinguish between multiple scenarios in the way the change of character of the equations occurs; these will be discussed in more detail in our companion paper \cite{Figueras:2026evh}. In this Letter, we only discuss the two cases directly relevant to the black string simulations:
\begin{itemize}
\item[(i)] \textit{Tricomi-type}: at least one eigenvalue ($\Delta_{-}$) goes to zero;  
\item[(ii)] \textit{Keldysh-type}: at least one eigenvalue ($\Delta_{+}$) diverges.
\end{itemize}
In both situations, the characteristic speed in some direction becomes imaginary. However, as noted in \cite{Figueras:2020dzx,Bernard:2019fjb}, in a numerical simulation it is not possible to actually see a dynamical change of character in a Keldysh-type transition because the system becomes extremely stiff as the characteristic speed blows up. Nevertheless, the generalized eigenvalues encode the asymptotic behaviour needed to distinguish between the two types of transition, while providing spatially gauge-invariant diagnostics. In addition, $\Delta_{\pm}$ vary continuously from $1$ in the GR limit to their asymptotic values. Thus, the behaviour of $\Delta_{\pm}$ also provides a convenient and quantitative measure of how much the EGB system deviates from GR.

In Fig.~\ref{fig:hyper} we show the time evolution of the global maximum of $\Delta_{+}$ (bottom) and the global minimum of $\Delta_{-}$ (top). The type of hyperbolicity loss depends on the sign of $\lambda^{\mathrm{GB}}$. For $\lambda^{\mathrm{GB}}>0$ the minimum of $\Delta_{-}$ declines steeply near the breakdown and approaches zero, signalling a Tricomi-type transition. 
For $\lambda^{\mathrm{GB}}<0$ the maximum of $\Delta_{+}$ grows by at least two orders of magnitude, indicating that the characteristic speed diverges and a Keldysh-type transition occurs. An analogous sign-dependence of the transition type appears in scalar \cite{Figueras:2025wtx} and scalar-tensor models \cite{Bernard:2019fjb}.
The spatial distribution of the transition is shown in Fig.~\ref{fig:AH_Delta}. For both signs, the change of character occurs just inside the AH, localised in the thinnest region of the string where the neck joins a blob \footnote{Recall that in our simulations we effectively turn off the Gauss-Bonnet term further inside the AH and hence the transition cannot occur too close to the centre of the string.}. This confirms that the loss of hyperbolicity is driven locally by the physical mode in the strongly coupled neck, and is therefore not an artefact of the mCCZ4 formulation or of the gauge choice.

\end{document}